\documentclass[reprint,prd,nofootinbib]{revtex4-1}
\usepackage{graphicx}
\usepackage{xcolor}
\usepackage{amsmath}
\usepackage[abbreviations]{siunitx}
\usepackage{subfigure}
\usepackage[hidelinks]{hyperref}
\def\equationautorefname#1#2\null{%
  Eq.\;(#2\null)%
}
\def\figureautorefname#1\null{%
  Fig.#1\null
}
\def\sectionautorefname#1\null{%
  Sec.#1\null
}

\begin{document}
\title{Charged charm stars}
\author{Victor P. Gon\c{c}alves}
\email{barros@ufpel.edu.br}
\affiliation{High and Medium Energy Group, Instituto de F\'isica e Matem\'atica, Universidade Federal de Pelotas \\
Caixa Postal 354, 96010-900, Pelotas, RS, Brazil.}
\author{Lucas Lazzari}
\email{lucas.lazzari@ufpel.edu.br}
\affiliation{High and Medium Energy Group, Instituto de F\'isica e Matem\'atica, Universidade Federal de Pelotas \\
Caixa Postal 354, 96010-900, Pelotas, RS, Brazil.}
\date{\today}
\begin{abstract}
 The study of the general properties and stability of charm stars with a net electric charge is performed within the MIT bag model framework. We consider two different models for the electric charge distribution and demostrate that both imply stellar configurations with larger masses and that satisfy the  equilibrium condition. The dynamical  stability against radial oscillations is investigated. Our results demonstrate that the eigenfrequencies are modified by the presence of a net electric charge, but the instability, previously demonstrated for the electrically neutral case, is also present in charged charm stars.
 \end{abstract}

\maketitle

\section{Introduction}
\label{sec:intro}
The description of the hadronic matter  present in the densest neutron stars remains a challenge for the theory of strong interactions -- the Quantum Chromodynamics (QCD) (For a recent review see, e.g. Ref. \cite{Baym}). Since the proposition that the ground state of strong interacting matter should be described in terms of quarks degrees of freedom ~\cite{bodmer1971,witten1984}, many studies suggested that the densest observed neutron stars could actually be quark stars~\cite{glendenning1996,weber1999}. Although the most known form of quark star is the one that contains roughly the same numbers of up, down and strange quarks, the so-called strange star~\cite{alcock1986,alcock1988,alford2005,kurkela2010,fraga2014,jimenez2019}, other forms of quark stars have been investigated~\cite{wang2019,kettner1995,jimenez2019b}. Quark stars containing only up and down quarks were not expected to exist, but recent studies suggest that the stability of dense quark matter in bulk is model dependent, implying that quark matter may not be strange~\cite{holdom2018}. In our study, we will assume the validity of Bodmer-Witten hypothesis as proposed in the MIT bag model framework~\cite{witten1984} and that the densest neutron stars contain strange quark matter (SQM). 

One important consequence of the Bodmer-Witten hypothesis is that for very large values of the energy density, charm quarks may be present in the star, constituting the so-called charm stars~\cite{kettner1995}. In the quark star scenario, charm stars present themselves as a last possibility because bottom and top quarks are too heavy to constitute any stable configuration~\cite{glendenning1996}. 
Electrically neutral charm stars were previously investigated by Kettner and collaborators~\cite{kettner1995} in the MIT bag model framework and more recently by Jim{\'e}nez and Fraga~\cite{jimenez2019b} using an approach based on dense and cold perturbative Quantum Chromodynamics (pQCD). One of our goals is to expand these previous studies for the case of charged charm stars. As demonstrated in Refs. ~\cite{negreiros2009,arbanil2015,lazzari2020}, for the case of charged strange stars, the presence of a net electric charge modifies the general properties of the star, implying e.g. heavier stars with a larger radii. In particular, the stability against radial oscillations is modified by the electric charge distribution within the star.
 The results presented in Refs. 
   ~\cite{kettner1995,jimenez2019b} pointed out that  neutral charm stars are unstable against radial oscillations. One important open question is if  such conclusion is also valid for the charged case.

In this paper, we will investigate for the first time the equilibrium and stability of charged charm stars. 
In particular, we will analyze the hydrostatic equilibrium for   different electric charge distributions inside the star as well as 
the impact of radial oscillations on the stability of charged charm stars. In our analysis we will assume, for simplicity, the MIT bag model equation of state (EoS) ~\cite{chodos1974}, which describes  the quark matter as a relativistic degenerate ideal Fermi gas of quarks and the confinement is taking into account by a bag pressure. As shown in Ref. ~\cite{lazzari2020}, the MIT bag model predictions are similar to those derived using a more realistic EoS.

This paper is organized as follows. In~\autoref{sec:form}, we will present a brief review of the formalism, presenting the stellar structure equations and the MIT bag model EoS for massive quarks and leptons. In addition, the chemical equilibrium for  quark matter containing the charm quark will be discussed and the models for the electric charge distribution are presented. In~\autoref{sec:res}, we will present our results  for the mass-radius profile and for the fundamental mode of oscillation of charged charm stars. The  equilibrium and stability  of charm stars will be discussed considering different values of the net electric charge. Finally, in~\autoref{sec:conc}, we will summarize our main conclusions.    

\section{Formalism}
\label{sec:form}

In order to determine the general properties of charged quark stars we have to solve the stellar structure equations obtained from the Einstein-Maxwell field equations, for a line element of the form $\mathrm{d}s^2 = e^{2\nu(r)}\mathrm{d}t^2 - e^{2\lambda(r)}\mathrm{d}r^2 - r^2(\mathrm{d}\theta^2 + \sin^2\theta\, \mathrm{d}\phi^2)\,$. These structure equations consist in a system of equations given by
\begin{align}
  \label{eq:TOV-q}
  \frac{\mathrm{d}q}{\mathrm{d}r} & {} = 4\pi r^2 \rho_e e^{\lambda} \,,\\
  \label{eq:TOV-m}
  \frac{\mathrm{d}m}{\mathrm{d}r} & {} = 4\pi r^2 \epsilon  + \frac{q}{r}\frac{\mathrm{d}q}{\mathrm{d}r} \,, \\
  \label{eq:TOV-p}
  \frac{\mathrm{d}P}{\mathrm{d}r} & {} = -(\epsilon + P)\left(4\pi r P + \frac{m}{r^2} - \frac{q^2}{r^3}\right)e^{2\lambda}   \nonumber \\
  & {} \quad + \frac{q}{4\pi r^4}\frac{\mathrm{d}q}{\mathrm{d}r} \,, \\
  \label{eq:TOV-nu}
  \frac{\mathrm{d}\nu}{\mathrm{d}r} & {} = -\frac{1}{\epsilon + P}\left(\frac{\mathrm{d}P}{\mathrm{d}r} - \frac{q}{4\pi r^4}\frac{\mathrm{d}q}{\mathrm{d}r}\right) \,,
\end{align}
where $\rho_e(r)$ is the electric charge density, $q(r)$ and $m(r)$ represent the charge and mass within radius $r$, respectively. The metric potential $\lambda(r)$ is of the Reisser-Nordstr\"om type. To solve this system of equations one has to use their respective boundary conditions (for more details, see~\cite{arbanil2015,lazzari2020}).     

The stability of a stellar system can be determined by solving the pulsation equation~\cite{chandrasekhar1964,kokkotas2001} alongside the stellar structure equations. The pulsation equation consists in a Sturm-Liouville eigenvalue problem
\begin{equation}
  \label{eq:stumliouville}
  \frac{\mathrm{d}}{\mathrm{d}r}\left[{\cal{P}} \frac{\mathrm{d}u}{\mathrm{d} r} \right] + [{\cal{Q}} + \omega^2{\cal{W}}]u = 0
\end{equation}
where $u$ is the renormalized displacement function. For a charged system, we have that~\cite{brillante2014}
\begin{align}
  \label{eq:pqw}
  \cal{P} = {} & e^{\lambda+3\nu}r^{-2}\gamma P\,, \\
  \cal{Q} = {} &  (\epsilon + P)r^{-2}e^{\lambda + 3\nu}\times \nonumber\\
  & {} \times\left[\nu'\left(\nu' - 4r^{-1}\right) - (8\pi P  + r^{-4}q^2)e^{2\lambda}\right] \,,\\
  \cal{W} = {} &  e^{3\lambda+\nu}r^{-2}(\epsilon + P)\,,
\end{align}
where $\gamma$ is the adiabatic index. The eigenfrequencies $\omega$ determine if a stellar configuration is or not stable. If $\omega$ is real ($\omega^2 \geq 0$) the star is stable. On the other hand, an imaginary eigenfrequency ($\omega^2 < 0$) leads to an exponential growth of $u$, resulting in unstable solutions. Since $Q$ is real, we have that $\omega^2_0 < \omega^2_1 < \cdots < \omega^2_n < \cdots$ thus, looking only at the sign of the fundamental eigenfrequency squared ($\omega^2_0$) we may determine the stability of the star.     

\begin{figure}[!t]
  \includegraphics[width=\columnwidth]{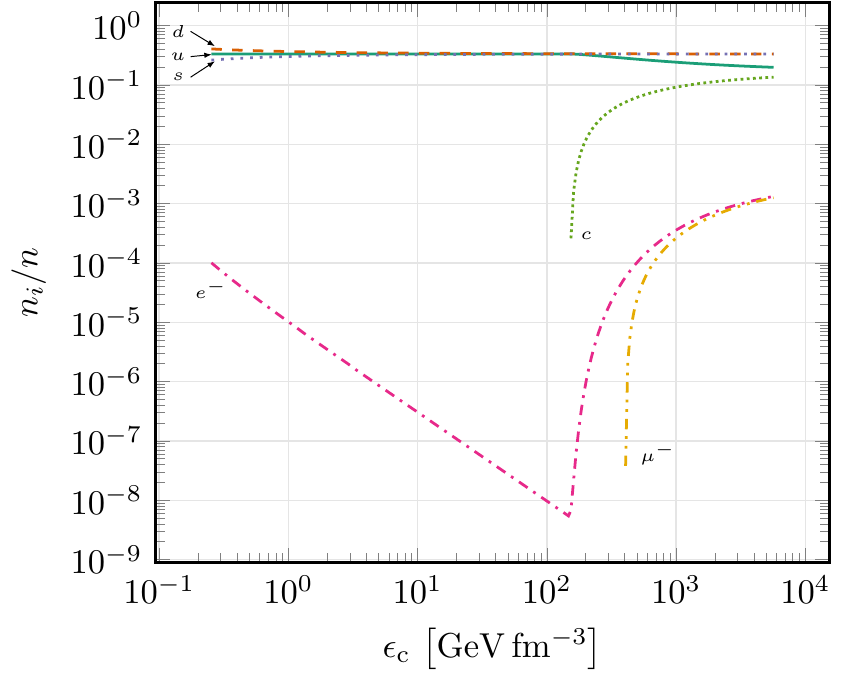}
  \caption{\label{fig:population} Particle number densities normalized by the total number density as a function of the central energy density.}
\end{figure}

\begin{figure*}[!t]
  \centering
  \subfigure{\includegraphics[width=.44\textwidth]{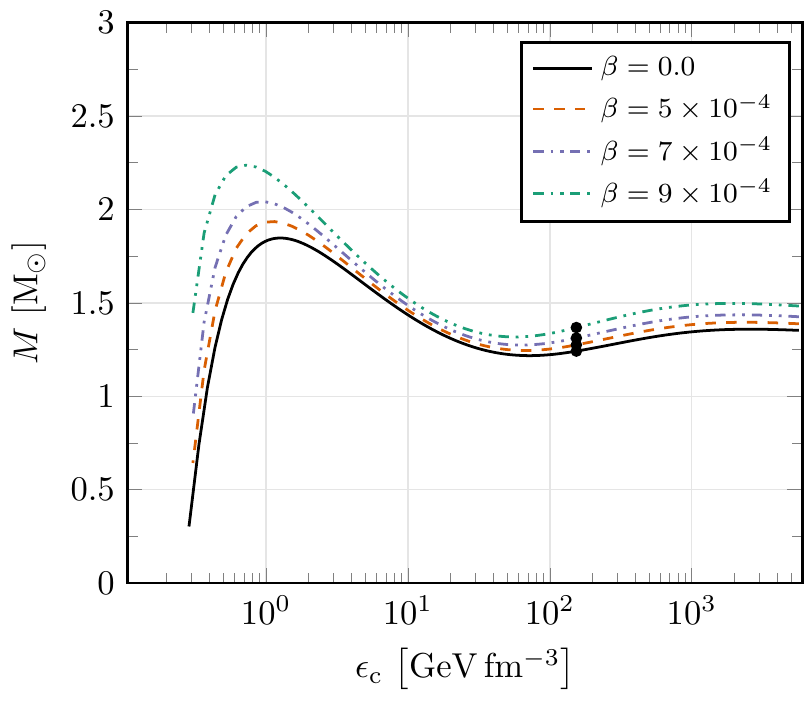}}\qquad
  \subfigure{\raisebox{2mm}{\includegraphics[width=.45\textwidth]{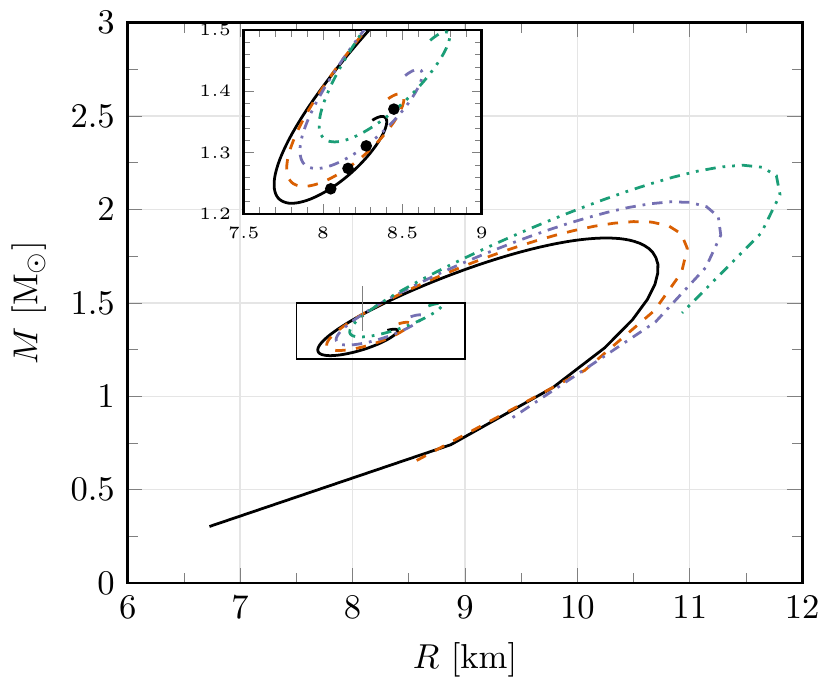}}}
  
  \caption{\label{fig:beta-MxecR} {\it Left panel:} Total gravitational mass as a function of the central energy density 
  for quark stars considering different values of $\beta$. The black filled circles in the left panel represent the  appearence of the charm quarks in the 
system. {\it Right panel:} Total gravitational mass as a function of the radius for quarks considering different values of $\beta$, with the black filled circles corresponding to possible charm star configurations. The predictions for $\beta = 0$ (black solid curve) correspond to electrically neutral quark stars.
  }
\end{figure*}
The solution of such system of equations depends on the EoS and on the electric charge distribution. Regarding the EoS, we chose to use the MIT bag model one, just for simplicity.  This EoS considers a Fermi gas of massive quarks and leptons, given by~\cite{glendenning1996}
\begin{align}
  \label{eq:MITp}
  p = & {} \sum_i\frac{g_i}{24\pi^2}\left[\mu_ik_i\left(\mu_i^2-\frac{5}{2}m_i^2\right) + \frac{3}{2}m_i^4\ln\left(\frac{\mu_i+k_i}{m_i}\right)\right] - \,, \nonumber \\
      & {} - B  \\
  \label{eq:MITe}
  \epsilon = & {}\sum_i\frac{g_i}{8\pi^2}\left[\mu_ik_i\left(\mu_i^2-\frac{1}{2}m_i^2\right) - \frac{1}{2}m_i^4\ln\left(\frac{\mu_i+k_i}{m_i}\right)\right] + \,, \nonumber \\
      & {} + B  \\
  \label{eq:MITn}
  n & {} = \sum_i n_i = \sum_i \frac{g_ik_i^3}{6\pi^2} \,.
\end{align}
where the index $i$ corresponds to the fermions species ($i = u, d, s, c, e^-, \mu^-$), $g_i$ is the degeneracy factor and $B$ is the bag pressure. The relation between the particles chemical potential and their momenta is simply $\mu_i^2 = k_i^2 + m_i^2$. Both pressure, energy density and particle number density are functions of the chemical potentials, which are related in order to satisfy chemical equilibrium. Considering that the quark star is on the final stages of its evolution and that the neutrinos have already left the star, we can say that the quark matter is in $\beta$-equilibrium, resulting in 
\begin{equation}
  \label{eq:chemicals}
  \mu_u = \mu_c ~,~ \mu_d = \mu_s ~,~ \mu_{e^-} = \mu_{\mu^-} ~,~ \mu_u = \mu_d + \mu_{e^-} \,.
\end{equation}
From the relations above, we see that from the six chemical potentials only two are independent variables, so we still require a last expression to relate them. The last condition to determine the chemical potentials is the global charge neutrality, expressed as $q = \displaystyle{\sum_i}q_i n_i = 0$, where $n_i$ corresponds to each summed term in~\autoref{eq:MITn} and $q_i$ are the respective  electric charges. From this, we numerically obtain $\mu_{e^-}$ with a root finding method while keeping $\mu_s$ as the independent variable. The relations between each particle species number density and the energy density are presented in~\autoref{fig:population}. We clearly see that the production of charm quarks starts on a very high energy density regime, which corresponds to a threshold of \SI{153.7}{GeV.fm^{-3}}. For  higher densities, the contribution of muons is non -- negligible and similar to the electron number density.

It is important to emphasize that the global neutrality condition does not imply in the absence of a net electric charge in the star. In fact, quark stars are believed to have a quark matter core with a surrounding electrosphere~\cite{alcock1986,negreiros2009} generating ultrahigh electric fields, of the order of~\SI{e21}{V.m^{-1}}. The electrosphere interacts with quark matter through electrostatic interaction and they are separated from each other by several hundred Fermi~\cite{alcock1986}. These ultrahigh electric fields generate a pressure~\cite{negreiros2009,arbanil2015} that grows with increasing charge throughout the star and reaches the same order of magnitude as the pressure from the quark gas near the stellar surface. The pressure from the electric field also has to be balanced by gravitational attraction, implying in stars with larger masses and radii. We will consider two types of charge distribution, as was done in Ref.~\cite{arbanil2015}. One is proportional to the third power of the radial coordinate, $q(r) = \beta r^3$, where $\beta \equiv Q/R^3$, and $Q$ and $R$ are the star's total charge and radius, respectively. We use $\beta$ in units of $\si{M_\odot.km^{-3}}$. Hereafter, we will denote  this distribution by  $\beta$-distribution. Moreover, we also will consider a charge density proportional to the energy density, i.e, $\rho_e = \alpha \epsilon$, where $\alpha$ is a dimensionless proportionality constant~\cite{arbanil2015,lazzari2020}, which will be denoted as $\alpha$-distribution. In the next section, we will present our predictions for the total gravitational mass as a function of the central energy density for charged quark stars considering different charge distributions. We also will estimate the eigenfrequencies from the radial oscillations modes and discuss the stability of charged charm stars.

\begin{figure*}[!t]
  \centering
  \subfigure{\includegraphics[width=.45\textwidth]{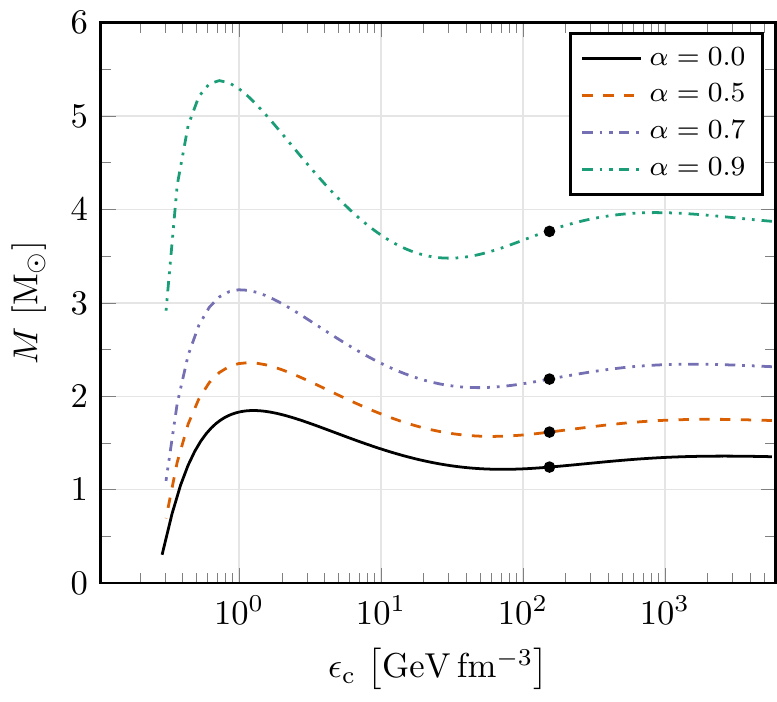}}\qquad
  \subfigure{\raisebox{1mm}{\includegraphics[width=.45\textwidth]{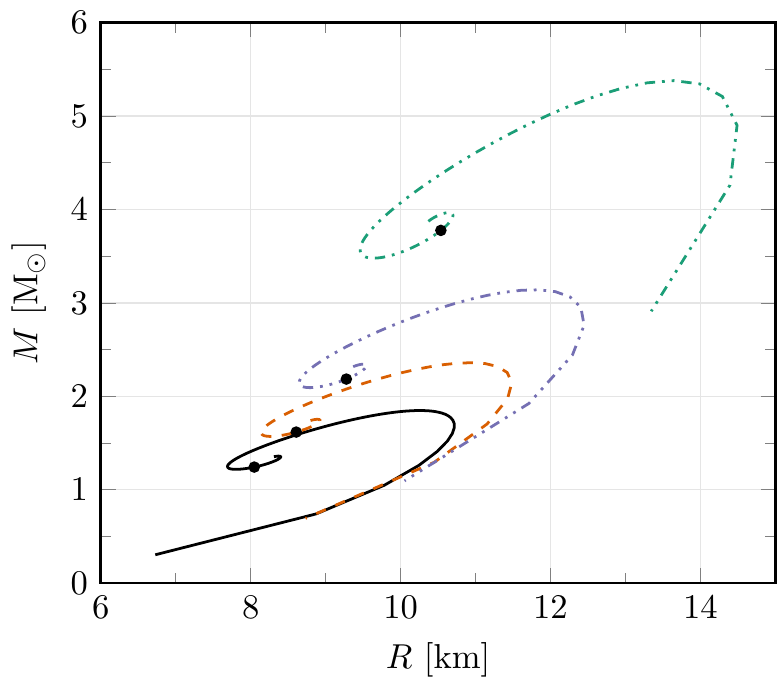}}}

  \caption{\label{fig:alpha-MxecR}  {\it Left panel:} Total gravitational mass as a function of the central energy density 
  for quark stars considering different values of $\alpha$. The black filled circles in the left panel represent the  appearence of the charm quarks in the 
system. {\it Right panel:} Total gravitational mass as a function of the radius for quarks considering different values of $\beta$, with the black filled circles corresponding to possible charm star configurations. The predictions for $\alpha = 0$ (black solid curve) correspond to electrically neutral quark stars.}
\end{figure*}

\section{Results}
\label{sec:res}
Initially, let's consider electrically neutral quark stars. Our predictions for the total gravitational mass as a function of the central energy density
and of the radius are represented by the black solid lines in ~\autoref{fig:beta-MxecR}. The results presented in the left panel indicate that the neutral quark stars have two branches of configurations in equilibrium that satisfy the condition $\partial M/\partial \epsilon_{\mathrm{c}} > 0$. The first one corresponds to the widely known strange stars and the second one corresponds to charm stars. The appearance of charm quarks in the system is represented by the black filled circles in the figure. The results presented in the right panel point out that the possible neutral charm star configurations are characterized by a mass of the order of 1.2 $M_{\odot}$ and a very small radius ($\approx 8$ km). 

In what follows we will investigate the impact of a charge distribution within the star in the basic properties of the charm star. The results for the 
the $\beta$-distribution, where $q(r) = \beta r^3$, are also presented in ~\autoref{fig:beta-MxecR} for different values of $\beta$. We have that 
the presence of charge increases the gravitational mass and the radius of the stellar configurations because of the high pressure due to the electric field near to the surface of the star. As in the neutral case, the appearance of charm quarks in the system occurs  for energy densities above \SI{153.7}{GeV.fm^{-3}}. Our results also indicate that the presence of a electric charge distribution also produces equilibrium configurations where $\partial M/\partial \epsilon_{\mathrm{c}} > 0$, with the associated radius increasing for larger values of $\beta$. Such behaviours are similar to those derived in Ref.~\cite{lazzari2020} for charged strange stars.

In~\autoref{fig:alpha-MxecR}, we present our predictions for the total gravitational mass as a function of the central energy density and the radius considering the $\alpha$-distribution, where $\rho_e = \alpha \epsilon$,  and assuming different values of $\alpha$.
Similarly to derived before, the presence of a charge distribution implies charm stars with larger masses and radii. As in the case of charged strange stars
~\cite{lazzari2020}, the predictions are strongly dependent of the value of $\alpha$. One important aspect is that equilibrium configurations are present for all values of $\alpha$.

\begin{figure}[!t]
  \includegraphics[width=\columnwidth]{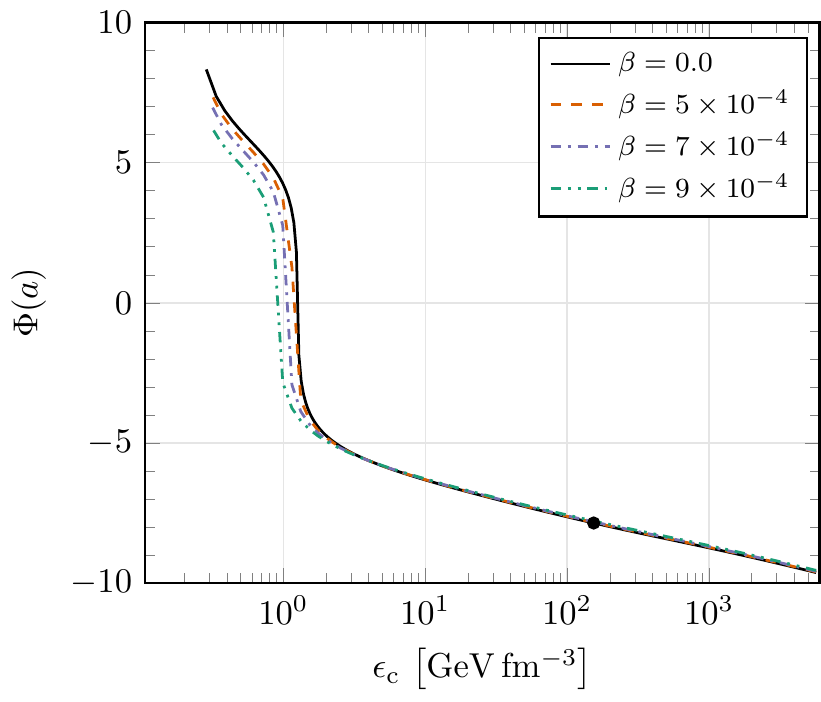}

  \caption{\label{fig:beta-fxec} Fundamental eigenfrequency vs central energy density for charged quark stars, using the $\beta$-distribution for different values of $\beta$. For convenience, we used $\Phi(a) = \mathrm{sign}(a)\ln(1 + |a|)$, where $a = (\omega_0/\si{kHz})^2$.}
\end{figure}

The results derived above indicate that equilibrum condition $\partial M/\partial \epsilon_{\mathrm{c}} > 0$ is also satisfied by charged charm stars. As discussed in previous studies ~\cite{kettner1995,jimenez2019b}, such condition is necessary but not sufficient to determine the stability of charm stars.
In order to establish the stability, it is fundamental to analyze the behaviour of the radial oscillation modes.  
In our study, the stability of charged charm stars will be investigated by the analysis of the sign of the fundamental eigenfrequency 
throught  the sign of the function $\Phi$, defined as $\Phi(a) = \mathrm{sign}(a)\ln(1+|a|)$, where $a = (\omega_0/\si{kHz})^2$. 
The results for the  $\beta$-distribution are presented in~\autoref{fig:beta-fxec}. We have that positive branch occurs for small central densities, where are the configurations for strange stars. On the other hand, for the energy densities associated to charm stars ($\epsilon_c \ge$ \SI{153.7}{GeV.fm^{-3}}), the fundamental eigenfrequency is negative, which implies that the charged and neutral charm stars are unstable against radial oscillations. Our results indicate that such conclusion is not sensitive to the value of $\beta$.

\begin{figure}[!t]
  \includegraphics[width=\columnwidth]{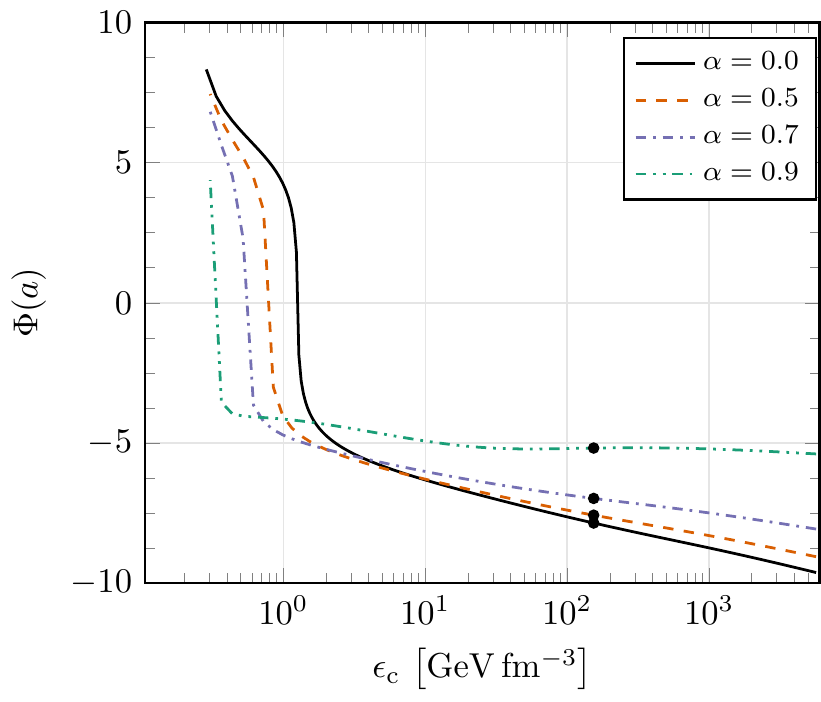}
  
  \caption{\label{fig:alpha-fxec} Fundamental eigenfrequency vs central energy density for charged quark stars, using the $\alpha$-distribution for different values of $\alpha$. For convenience, we used $\Phi(a) = \mathrm{sign}(a)\ln(1 + |a|)$, where $a = (\omega_0/\si{kHz})^2$.}
\end{figure}

The predictions for  the $\alpha$ - distribution are presented  in~\autoref{fig:alpha-fxec}. For this distribution we also have that for the energy densities associated to neutral and charged charm stars, the fundamental eigenfrequencies are negative. In contrast to the $\beta$ - distribution, the predictions are dependent on the value of $\alpha$, but the sign of  $\Phi(a)$ is not changed by increasing $\alpha$. 
Therefore, charm stars are not stable in Nature, independently from the presence of a charge distribution. An important open question is if charm stars could exist for a brief period of time, say, after the collision of two neutron stars or strange stars.

\section{Summary}
\label{sec:conc}
In this paper, we have investigated, for the first time, the general properties of electrically charged charm stars within the framework of the MIT bag model using different electric charge distributions. We have shown that a net electric charge in charm stars leads to stellar configurations with larger masses and radii. We have shown that the presence of a net electric charge implies in charm stars that satisfy the equilibrium condition, $\partial M/\partial \epsilon_{\mathrm{c}} > 0$,  as previously derived for neutral charm stars. We demonstrated that charged charm stars are unstable against radial oscillations. Our results indicate that  charm stars cannot exist as stable systems in Nature, regardless of their electric charge.      

\section*{Acknowledgements}
This work was partially financed by the Brazilian funding agencies CNPq, Coordena\c{c}\~ao de Aperfei\c{c}oamento de Pessoal de N\'ivel Superior (CAPES) -- Finance Code 001, FAPERGS and INCT-FNA (process number 464898/2014-5).


\end{document}